\documentclass[prd,aps,superscriptaddress,amsmath,amssymb,amsfonts,floatfix]{revtex4-2}
\usepackage{amssymb,amsmath,amsthm,amsbsy,color,graphicx,times}
\pdfoutput=1
\usepackage[ansinew]{inputenc}
\usepackage[english]{babel}
\usepackage{color}
\usepackage{braket}
\usepackage{booktabs}
\usepackage{tabularx}
\usepackage{array}
\usepackage{bbold}
\usepackage{natbib}

\begin{document}

\title{ On the geometric phase   and Majorana neutrinos }

\author{A. Capolupo}
\affiliation{Dipartimento di Fisica "E.R. Caianiello" Universit\'a di Salerno,  and INFN - Gruppo Collegato di Salerno, Italy}
\author{S.M. Giampaolo}
\affiliation{International Institute of Physics, Universidade Federal do Rio Grande do Norte, 59078-400 Natal-RN, Brazil}
\author{B. C. Hiesmayr}
\affiliation{Faculty of Physics, University of Vienna, W\"{a}hringerstrasse 17, 1090 Vienna, Austria.}
\author{G. Lambiase}
\affiliation{Dipartimento di Fisica "E.R. Caianiello" Universit\'a di Salerno,  and INFN - Gruppo Collegato di Salerno, Italy}
\author{A. Quaranta}
\affiliation{Dipartimento di Fisica "E.R. Caianiello" Universit\'a di Salerno,  and INFN - Gruppo Collegato di Salerno, Italy}

\date{\today}

\begin{abstract}

We analyze the geometric phase for neutrinos and we demonstrate that the geometric invariants associated to transitions between different neutrino flavors, for Majorana neutrinos, can depend on the representation of the mixing matrix and then are sensitive to the nature of neutrinos. The dependence of geometric invariants on the Majorana phase cannot be eliminated by a charged lepton rephasing transformation. By considering kinematic and geometric approach we also demonstrate that the Majorana phase is relevant in the projective Hilbert space. Geometric invariants can in principle be used as tools to distinguish between Dirac and Majorana neutrinos. 
\end{abstract}

\maketitle

\section{Introduction}

The mixing and oscillation of particles represents a particularly interesting phenomenon in modern physics, both from the experimental and the theoretical point of view, and its connections with cosmology and astrophysics are extensively studied \cite{Endoh2002,Moffat2019,Azari2020,Stirner2018,CapolupoCurv,CapolupoDens,CapolupoFlavor}. Many experiments have shown the existence of neutrino oscillations and as a consequence that neutrinos are massive particles \cite{Nakamura1,Nakamura2,Nakamura3,Nakamura4,Nakamura5,Nakamura6}. One of the open problems of neutrino physics is the nature of neutrinos. In fact, they can be Majorana fermions (in which case the particle state coincides with the antiparticle state), or Dirac fermions (in which case the particle state is distinct from the antiparticle state). The detection of neutrinoless double beta decay could indicate the nature of neutrinos  \cite{Giuliani}, as Majorana fermions allow phenomena in which there is no conservation of the total lepton number.
However, such a decay is very rare and has not been observed so far. Therefore, the analysis of physical quantities showing differences between Majorana and Dirac fermions is extremely important.
It  has been shown that the geometric phases for neutrinos ~\cite{Capolupo2018} and the oscillation formulas in the presence of decoherence ~\cite{Capolupo2019} can be influenced by the Majorana CP-violating phase. Phases and, particularly, geometric phases, are often analyzed to gain information on fundamental particles and interactions ~\cite{Phases1,Phases2,Phases3,Phases4}

In the present paper we discuss several aspects related to the Majorana CP-violating phase in neutrino oscillations. We analyze the impact and the appearance of the Majorana phase in various observables, such as geometric invariants, and the entanglement related to neutrino interactions. We argue that the dependence of these quantities on the Majorana phase cannot be eliminated by means of a charged lepton field rephasing in the charged current Lagrangian, since the Majorana phase cannot be removed from the mixing matrix. We discuss the conceptual issues and limits of the redefinitions of the neutrino mixing matrix based on the charged lepton field rephasing. In particular we show that different parametrizations of the mixing matrix for Majorana neutrinos lead to flavor mixing Hamiltonians which are in principle distinct.

We argue that the transformations relating different parametrizations of the mixing matrix cannot be interpreted as a change of basis if the neutrinos are Majorana, but only if they are of Dirac type.

We explore the possibility of highlighting the difference between Majorana neutrinos and Dirac neutrinos through the geometric phase, and 
discuss in detail their definition. We introduce noncyclic geometric invariants which are $U(1)$ gauge and reparametrization invariant and show that any term proportional to the identity in the Wolfenstein effective Hamiltonian has no effect on them. We consider distinct parametrizations of the neutrino mixing matrix (in the two flavor case for simplicity) and we show that the geometric invariants depend on the parametrization chosen. In particular, they depend or not depend on the Majorana phase according to the parametrization employed. We prove that the appearance of the Majorana phase cannot be ascribed to a lepton field rephasing transformation.

We show that using the kinematic and geometric approaches, one can define  geometric invariants that bear an explicit dependence on the Majorana phase. Therefore, the geometric invariants defined in the present paper show a distinction between Dirac and Majorana neutrinos, since they depend on the CP-violating Majorana phase.
The question on how experiments could ever  detect the geometric phase of neutrinos remains open. However the analysis of the  geometric phase could open a new scenario to investigate the neutrino properties. Further comments about the role of the wavefunction collapse on the possible measurement of the geometric phase are reported in the appendix A.

The paper is organized as follows. In Section II, we show that, for Majorana neutrinos, it is not possible to get rid of the Majorana phase
by appropriately choosing a rephasing of  the charged lepton field in the charge current Lagrangian. Moreover, in section III we describe a gedanken experiment  showing that the redefinition of the mixing matrix, due to a
rephasing,   modifies some physical quantities.
In Section IV, we define two geometric invariants depending on the Majorana CP-violating phase which are in principle measurable. In Section V, we consider the kinematic and the geometric approach to the geometric phase. In the kinematic approach, we show that by using the matrix $U^{(1)}$ of eq. \eqref{MixingMatrices}, the total phase does not depend on the Majorana phase, while for $U^{(2)}$, this dependence is explicit.
Analyzing the geometric approach, we demonstrate that the Majorana phases  are geometrically relevant in the Projective Hilbert space and then that they
can affect geometric invariants.
In Section VI, we draw our conclusions. In the appendix we study the geometric phase and the collapse of the wave function as introduced in  and we show that the conclusions
of Ref.~\cite{Lu2021} are inconsistent and flawed.

\section{The role of Majorana phase in neutrino mixing}

We here consider the role of the Majorana phase and the parametrization of the neutrino mixing matrix in physical observable quantities. 
It is know that \cite{Giunti} the standard oscillation formulae are independent on the Majorana phase.
On the other hand, quantities like the neutrinoless beta decay rate \cite{PDG2020}, the oscillation formulas in presence of decoherence \cite{Capolupo2019} and the formulas for the hypothetical neutrino--antineutrino oscillations \cite{NeutrinoAntineutrino} depend explicitly on the Majorana phase. These quantities do not, however, depend on the parametrization of the mixing matrix for Majorana neutrinos.

In the following we will analyze the possibility of the existence of physical quantities that, apart from depending on the Majorana phase, show a dependence on the parametrization of the neutrino mixing matrix. The reason for doing so is that the reparametrization of the Majorana neutrino mixing matrix is obtained by means of charged lepton field rephasing and not by means of a proper change of basis within the neutrino Hilbert space. In this section we discuss why it cannot be considered as a proper change of basis and why this may potentially cause troubles.
In addition, in the next sections, we address situations in which both the Majorana phase and the parametrization of the mixing matrix are physically relevant.

We focus on two--flavor mixing for simplicity, but the analysis can be readily extended to three flavors. The neutrino mass Hamiltonian for the propagation in the vacuum, but similar considerations apply to the propagation in matter, can be written, when expressed on the flavor basis, in either of the two ways
  \begin{equation}
   H^{(1)} = \begin{pmatrix} m_{ee} & m_{e\mu} \\ m_{e\mu} & m_{\mu \mu}\end{pmatrix} \ \ \ \ \ \ \ \ \ H^{(2)} = \begin{pmatrix} m_{ee} & m_{e\mu} e^{i \phi} \\ m_{e\mu} e^{-i \phi} & m_{\mu \mu}\end{pmatrix} \ .
\end{equation}
It is fundamental to realize that the two Hamiltonians are equivalent if and only if the neutrinos are of the Dirac type.
Indeed, if the neutrinos are Dirac fermions, the phase $\phi$ in $H^{(2)}$ can be reabsorbed using a simple rephasing of the neutrino fields, under which the Dirac Lagrangian is invariant.
On the contrary, the Majorana Lagrangian is not invariant under global $U(1)$ transformations \cite{GiuntiKim2007}, and this fact implies that the phase freedom for the neutrino fields is lost.
That the two Hamiltonians $H^{(1)}, H^{(2)}$ are not equivalent in the Majorana case is also evident in their diagonalization, indeed the two Hamiltonians are diagonalized by two distinct unitary matrices
\begin{equation}\label{MixingMatrices}
  U^{(1)} = \begin{pmatrix} \cos \theta & \sin \theta e^{i \phi} \\- \sin \theta & \cos \theta e^{i \phi}\end{pmatrix} \ \ \ \ \ \ \ \ \ U^{(2)} = \begin{pmatrix} \cos \theta & \sin \theta e^{i \phi} \\- \sin \theta e^{- i \phi} & \cos \theta \end{pmatrix} \ ,
\end{equation}
respectively for $H^{(1)}$ and $H^{(2)}$.
It is easy to check that $H^{(1)}$ cannot be diagonalized by $U^{(2)}$ and $H^{(2)}$ cannot be diagonalized by $U^{(1)}$. Both mixing matrices bring the respective Hamiltonians to the same diagonal matrix according to
\begin{equation}\label{Diagonalization}
 (U^{(1)})^{\dagger} H^{(1)} U^{(1)} = H_{m}^{(1)} = \begin{pmatrix} m_1 & 0 \\ 0 & m_2 \end{pmatrix}= H_{m}^{(2)} = (U^{(2)})^{\dagger} H^{(2)} U^{(2)}  \ .
\end{equation}
It is the above equations that define the mixing matrix: if $H_f$ is the mass Hamiltonian in the flavor basis, all the possible mixing matrices $U$ are only those which satisfy $U^{\dagger}H_f U = H_m$.
Even if the matrices in the mass basis coincide ($H_{m}^{(1)}\equiv H_{m}^{(2)}$), the different mixing matrices define two distinct and non-equivalent sets of linear relations between the flavor states and the mass states.

Nevertheless, in the ref.~\cite{Giunti} the authors argued that one can freely move from the mixing matrix $U^{(1)}$ to the mixing matrix $U^{(2)}$ by rephasing the charged lepton fields in the weak charged current interaction:
\begin{equation}\label{ChargedCurrent}
 \mathcal{L}_{CC} = \frac{g}{\sqrt{2}} \sum_{a= e, \mu } \sum_{k = 1,2} \overline{a_L} (x) \gamma^{\rho} U_{a k} \nu_{kL} (x) W_{\rho} (x)   + h. c.
\end{equation}
where $U_{ak}$ denotes the neutrino mixing matrix, $W_{\rho}$ is the charged weak gauge field and the subscript $L$ is used to denote the left-handed fermion fields.

In the case in which the neutrino fields are of Majorana type, one is not free to rephase $\nu_{kL}$, however, of course, it is always possible to rephase the charged lepton fields $a_{L}$ which are of Dirac type.
However, in the ref.~\cite{Giunti} it is taken one step forward, i. e. the charged lepton field rephasing $a_L \rightarrow a_L e^{- i \phi_a}$ for $a= e,\mu$, is interpreted as defining a new mixing matrix
\begin{equation}\label{Rephasing}
 \tilde{U}_{ak} = e^{i \phi_a} U_{ak}  \ ,
\end{equation}
and clearly one can always choose the phases $\phi_{a}$ in order to bring $\tilde{U}$ either in the form $U^{(1)}$ or $U^{(2)}$.

If the neutrino mixing matrix is originally to be defined, from Eq. \eqref{Diagonalization}, then such an interpretation is not justified, because, as shown in Eq. \eqref{Diagonalization}, it leads to two distinct flavor mixing Hamiltonians. If one considers the equation \eqref{ChargedCurrent} as \emph{defining} the neutrino flavor states and thus the neutrino mixing matrix $U_{ak}$, then $U_{ak}$ inherits the rephasing freedom from the charged leptons. Also in this case, however, distinct mixing matrices lead to distinct neutrino mixing Hamiltonians.

The rephasing of $U_{ak}$ is not legitimate for at least three reasons:

\begin{enumerate}

\item The rephasing of $U_{ak}$ has consequences on the flavor Hamiltonian. Independently on whether the mixing matrix is defined by equation \eqref{Diagonalization} or equation \eqref{ChargedCurrent}, distinct mixing matrices $U^{(1)}$ and $U^{(2)}$ are associated to flavor Hamiltonians that are in principle distinct. Therefore the transformation of the mixing matrix $U$ by rephasing the charged lepton fields $e_L$ and $\mu_L$ cannot be done arbitrarily, since $U^{(1)}$ and $U^{(2)}$ are not physically equivalent relatively to the neutrino propagation.

The rephasing is referred to the weak charged current interaction, characterizing the processes of detection and creation of neutrinos. On the other hand the mixing and the mixing matrix are related to an interaction which is independent of the weak interaction and that characterizes the neutrino oscillations during its propagation. We point out that the weak interaction and the neutrino mixing are two different kinds of interactions and therefore it is completely arbitrary to modify the neutrino mixing matrix by rephasing charged leptons, which do not take part in the mixing phenomenon.

In other words, this interpretation would be legitimate if the neutrinos only participated in the weak interaction, evidently bringing no actual modification to the neutrino physics.
However, neutrinos also oscillate according to the flavor Hamiltonian $H_f$ and it is not acceptable that the (unphysical) rephasing of the charged lepton fields brings along a change in the neutrino flavor Hamiltonian.
Obviously, according to whether the mixing matrix is chosen as $U^{(1)}$ or $U^{(2)}$, one obtains respectively $H_f = H^{(1)}$ and $H_f = H^{(2)}$ which are in principle distinct.

\item Besides the impossibility to redefine the mixing matrix through the rephasing of the charged lepton fields, we also remark that no rephasing of the Majorana mass fields is allowed, as the Majorana Lagrangian lacks the necessary $U (1)$ invariance. Therefore, even if
 the matrices $U^{(1)}$ and $U^{(2)}$ are related by the unitary matrix of a change of basis, given by
\begin{equation}
 U = \begin{pmatrix} 1 & 0 \\ 0 & e^{- i \phi}  \end{pmatrix} \ ,
\end{equation}
this is not a legitimate change of basis if the neutrinos are Majorana, but only if the neutrinos are of Dirac type.

We conclude that if the mixing matrix is to be defined from the flavor Hamiltonian via Eq.~\eqref{Diagonalization}, then the identification in Eq.~\eqref{Rephasing} is not legitimate, unless the new mixing matrix diagonalizes the same flavor Hamiltonian (which is not the case, of course, for $U^{(1)}$ and $U^{(2)}$).
This argument is independent of whether the two Hamiltonians $H^{(1)}$ and $H^{(2)}$ produce distinct observable effects or not.

\end{enumerate}

Let us also remark that while the standard oscillation formulae are not affected by the Majorana phase $\phi$~\cite{Giunti}, the latter may well show up in other circumstances, for instance in the propagation in a dissipative medium~\cite{Capolupo2019} for which the oscillation formulas depend on the Majorana phase.
Finally, it is not at all clear why the charged lepton field rephasing should have any effect on neutrinos, which should not be affected by the latter, because then it is obvious that any quantity depending only on the neutrino oscillatory evolution suffers no modification from the charged lepton field rephasing.

\section{Majorana phase: a gedanken experiment}

In this section we show that the charged lepton field rephasing of Eq. \eqref{Rephasing} can have consequences, not only on neutrino mixing, but also in hypothetical neutrino interactions. In the following we consider the interaction of neutrinos with an auxiliary $2$-level system (the auxiliary system would be called ``ancilla'' in the context of quantum information). The interaction has the effect of transferring part of the information about the neutrino state to the state of the auxiliary system.
We describe a gedanken experiment to show that the redefinition of the mixing matrix due to the charged lepton field rephasing in Eq.~\eqref{Rephasing} can have consequences on physical quantities that are in principle measurable.
To show this fact let us start by parametrizing the two-flavor mixing matrix, by means of the parameter $\psi$, as
\begin{equation}
 U(\psi) = \begin{pmatrix} \cos \theta & \sin \theta e^{i \phi} \\ -\sin \theta e^{i\left(\psi - \phi \right)} & \cos \theta e^{i \psi} \end{pmatrix} \ .
\end{equation}
Notice that $U(\psi)$ interpolates between $U^{(1)} = U(\phi)$ and $U^{(2)} = U (0)$.

As it is well known, when a neutrino is emitted , say at $t=0$, its state, depending on the particular process, coincides with a definite flavor state. Then the neutrino undergoes flavor oscillations. Therefore, at later times $t > 0$ the neutrino state will be, in general, described by a state which is a linear combination of the two flavor states
\begin{eqnarray}
 \nonumber \ket{\nu (t)} &=& \alpha (t) \ket{\nu_e} + \beta (t) \ket{\nu_{\mu}} \nonumber \\
 \nonumber &=& \alpha(t) \left( \cos \theta \ket{\nu_1} + \sin \theta e^{i \phi} \ket{\nu_2} \right) + \beta (t) \left( - \sin \theta e^{i \left(\psi - \phi \right)}\ket{\nu_1} + \cos \theta e^{i \psi} \ket{\nu_2} \right) \\
 &=& \left(\alpha (t) \cos \theta - \beta(t) \sin \theta e^{i \left(\psi - \phi \right)} \right) \ket{\nu_1} + \left( \beta (t) \cos \theta e^{i \psi}  + \alpha (t)\sin \theta e^{i \phi}\right) \ket{\nu_2} \ ,
\end{eqnarray}
with $|\alpha(t)|^2 + |\beta (t)|^2 = 1$.
Suppose now that at a time $t_1 > 0$ an interaction with an auxiliary $2$-level system is switched on.
The precise nature of the auxiliary system (atoms, photons or else) is of no interest for our considerations.
Let us denote the basis states of the auxiliary system as $\ket{0}, \ket{1}$, and assume that its initial state is $\ket{A(t_1)} = \frac{1}{\sqrt{2}}\left(\ket{0} + \ket{1} \right)$.
Let us consider that the interaction Hamiltonian, written on the basis $\lbrace \ket{\nu_1}, \ket{\nu_2} \rbrace \otimes \lbrace \ket{0} \ket{1} \rbrace$ has the form
\begin{equation}\label{AuxiliaryHamiltonian}
 H = \omega \begin{pmatrix} 1 & 0 & 0 & 0 \\ 0 & -1 & 0 & 0 \\ 0 & 0 & -1 & 0 \\ 0 & 0 & 0 & 1 \end{pmatrix}
\end{equation}
with $\omega$ a constant with dimensions of energy.
Also in this case the nature of the interaction is of no interest for our analysis.
For the sake of simplicity, we assume that the duration of the interaction $\tau$ is much smaller than the neutrino oscillation period $T=\frac{8\pi E}{\Delta m^2}$, so that during the interaction process we can neglect the oscillations.
This assumption allows us to consider the two complex parameters that describe the linear superposition, i.e. $\alpha(t_1)$ and $\beta(t_1)$, as constant for the duration of the interaction with the auxiliary system.
This hypothesis is not crucial, but has the advantage of making the analysis easier without affecting the generality of the results.

Since the Hamiltonian $H$ of Eq.~\eqref{AuxiliaryHamiltonian} cannot be reduced to the sum of local terms, each acting on a single subsystem, it develops a non-vanishing entanglement between the neutrino and the auxiliary system. Effectively this allows the information about the quantum state to flow from the first to the second system.
Indeed, letting the full system evolve under the action of $H$ for a time $\tau$, and tracing out the neutrino degrees of freedom at the end of the interaction time $t_2 = t_1 + \tau$, one finds that the auxiliary system is in a mixed state with density matrix
\begin{equation}
\label{ancilla_t2}
 \rho_A (t_2) = \frac{1}{2} \begin{pmatrix} 1 & \gamma (t_2, t_1) \\ \gamma^* (t_2 , t_1) & 1 \end{pmatrix}
\end{equation}
where
\begin{eqnarray}
\label{gamma}
\nonumber \gamma (t_2,t_1) &=& -e^{- i \left( 2 \omega \tau + \psi + \phi \right)} \Bigg[\frac{e^{4i \omega \tau}-1}{2} \sin 2 \theta \left( e^{2 i \psi} \beta(t_1) \alpha^* (t_1)+e^{2i \phi}\beta^{*}(t_1) \alpha (t_1)\right) \\
&-& \frac{e^{i(\psi + \phi)}}{2} \left(1 + e^{4 i \omega \tau} + \cos 2 \theta \left(1-e^{-4i\omega \tau} \right)\left(|\beta(t_1)|^2 - |\alpha(t_1)|^2 \right) \right)\Bigg] \ .
\end{eqnarray}
As it can be seen, at $t=t_2$ the off diagonal elements of the reduced density matrix of the auxiliary system depend on $\psi$.  It is worth noting that this analysis can be generalized to the case of many interactions with the auxiliary system. If the auxiliary system interacts consecutively with $n$ identical neutrinos, the $\gamma(t_2,t_1)$ is replaced by
\begin{equation}\label{NconsecutiveInteractions}
 \gamma_{n} (t_2,t_1) = \left( \gamma(t_2,t_1)\right)^n
\end{equation}

The value of $\psi$ can be extracted observing several different quantities.

To provide an example,  we can consider the 2-Renyi entropy of $\rho_A(t_2)$ that is a measure of the entanglement developed between the neutrino and the auxiliary system and~\cite{Nielsen2000, Horodecki2002, Plenio2007, Giampaolo2013}, that has the advantage, with respect to the other entropy-based entanglement measures to be associated, at least in some experimental devices, to experimentally accessible quantities~\cite{Bovino2006, Brydges2019, Lesche2004, Islam2015, Abanin2012}.
The 2-Renyi entropy is defined as $S_2 = -\ln(P(\rho_A(t_2)))$,  and $P(\rho_A(t_2)) = Tr \left((\rho_A(t_2))^2\right)$ is the purity of $\rho_A(t_2)$.
In our system the 2-Renyi entropy becomes
\begin{equation}\label{2REntropy}
 S_2=-\ln\left(\frac{1}{2}+|\gamma|^2\right)
\end{equation}
A typical behavior of $S_2$ as a function of $\psi$ is shown in the figure (1) for sample values of the mixing angle $\theta = \frac{\pi}{4}$, Majorana phase $\phi = \frac{\pi}{10}$, $\alpha(t_1)=\frac{1}{2} e^{\imath \pi/4}$, $\beta(t_1)=\frac{\sqrt{3}}{2}  e^{-\imath 3\pi/8}$ and $\omega \tau = 0.4$  for several values of $n$.

\begin{figure}[t!]
	\includegraphics[width=0.5\columnwidth]{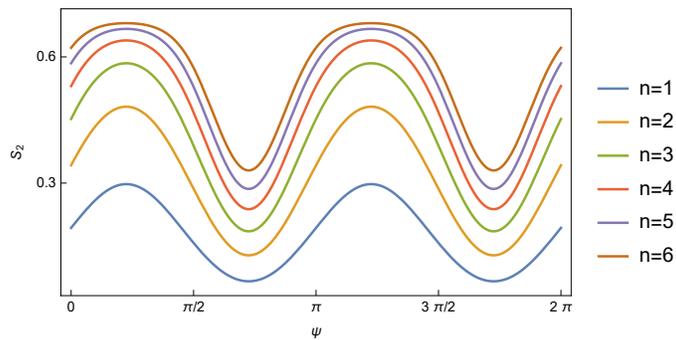}
\caption{Behavior of $S_2$ (eq. \eqref{2REntropy}) as function of $\psi$ assuming: $\theta=\frac{\pi}{4}$, $\phi=\frac{\pi}{10}$, $\alpha(t_1)=\frac{1}{2} e^{\imath \pi/4}$, $\beta(t_1)=\frac{\sqrt{3}}{2}  e^{-\imath 3\pi/8}$ and $\omega \tau = 0.4$  for several values of $n$.}\label{fig_1}
\end{figure}

Before going further, let us briefly comment on the results that we have presented in this section.

The experiment here proposed, most likely, cannot be realized with the current technologies, nor with those of the near future. In order to extract information on the neutrino state, we are considering the interaction of a neutrino with an auxiliary system that can be easily controlled, and hence must be an atom or at most a massive particle as a neutron or a proton. For instance, a two level system can be generated by considering spatially separated superpositions like the ones suggested in BMV ( Vacuum magnetic birefringence) experiments \cite{Bose}. In this setting the gravitational interaction between the ancillary system and the neutrino naturally provide an interaction as the one analyzed in this section \cite{Giampaolo2019}.

Setting up an interaction of the kind \eqref{AuxiliaryHamiltonian}, with the degree of control needed the perform an entanglement measurement is certainly prohibitive for the current technology.
However, the proposed setup makes evident how it is possible, at least in principle, to design experiments in which one can observe and determine the value of the global phase that multiplies one of the two flavor states, due to its effect on physical quantities such as the $2$-Renyi entropy. Consequently, at least in principle, an experiment of this kind is able to discern the actual mixing matrix.

\section{Geometric invariants and Majorana phase}

In Refs.~\cite{Lu2021,Johns2021} it is argued that the appearance of the Majorana phase in the expression of the geometric phase is a consequence of an unphysical rephasing transformation.
In other words, the authors affirmed that by appropriately choosing a rephasing transformation, it is possible to get rid of the Majorana phase.
This argument of rephasing freedom is not new and it is essentially borrowed from~\cite{Giunti}.
In the following we analyze the noncyclic geometric invariants associated to neutrino mixing. First we discuss their $U(1)$ gauge invariance and then we show how they can depend on the Majorana phase.

Let us analyze these points in detail.
The geometric phases at a time $t \sim z>0$ for an initial electron neutrino state $\ket{\nu_{e}}$ and that evolves under the action of the mass Hamiltonian, is given by
\begin{eqnarray}\label{fase1}
\nonumber \Phi^{g }_{\nu_{e}}(z)  & = & \Phi^{tot}_{\nu_{e}}(z) - \Phi^{dyn}_{\nu_{e}}(z)
\\
 & = &   \arg \left[\braket{\nu_{e}(0)| \nu_{e}(z)}\right] - \Im \int_{0}^{z}
 \braket{\nu_{e}(z^{\prime})| \dot{\nu}_{e}(z^{\prime})}  d z^{\prime}\,,
 \\ \nonumber
  & = & \arg \left[ \cos \left(\frac{\Delta m_{m}^{2} z}{4 E}\right) + i \cos 2\theta_{m} \sin \left(\frac{\Delta m_{m}^{2} z}{4 E}\right) \right]
-\frac{\Delta m_{m}^{2} z}{4 E} \,\cos 2\theta_{m}  \,.
\end{eqnarray}
A similar result is obtained for an inital muon neutrino state $\ket{\nu_{\mu}}$, indeed one has  $\Phi^{g }_{\nu_{\mu}}(z) = - \Phi^{g }_{\nu_{e}}(z)$.
Both $\Phi^{g }_{\nu_{\mu}}(z)$ and $\Phi^{g }_{\nu_{e}}(z)$ are gauge invariant and re-parametrization invariant.

In deriving the explicit expression of $\Phi^{g }_{\nu_{\mu}}(z)$ in eq.~\eqref{fase1}, we have neglected a term proportional to the identity in the neutrino Hamiltonian $H_0 = \left( \frac{m_1^2 + m_2^2}{4E} + \frac{G_F n_e}{\sqrt{2}} \right) \mathbb{I}$.
It is straightforward to show that such a term has no effect whatsoever on the previously defined invariants.
The inclusion of this term, (a term proportional to the identity matrix), amounts to an additional time-dependent global phase factor $\ket{\nu_{e,\mu}(z)} \rightarrow e^{i \lambda (z)} \ket{\nu_{e,\mu}(z)}$, with $\lambda (z) = - \int_{0}^{z} dz' \left( \frac{m_1^2 + m_2^2}{4E} + \frac{G_F n_e}{\sqrt{2}} \right) $.
Computing the invariants taking into account the additional phase factor, one finds
\begin{eqnarray}
\label{fase2}
 \nonumber \Phi^{g }_{\nu_{e}}(z)  &=& \arg \left[\langle \nu_{e}(0)| \nu_{e}(z)\rangle  e^{i \left(\lambda(z) - \lambda(0) \right)}\right] - \Im \int_{0}^{z}  \langle  \nu_{e}(z^{\prime})| \dot{\nu}_{e}(z^{\prime})\rangle  d z^{\prime} - \Im i \int_{0}^{z} \lambda' (z') dz' \\
 \nonumber &=&  \arg \left[\langle \nu_{e}(0)| \nu_{e}(z)\rangle  \right] + (\lambda (z) - \lambda (0)) - \Im \int_{0}^{z}  \langle  \nu_{e}(z^{\prime})| \dot{\nu}_{e}(z^{\prime})\rangle  d z^{\prime} - (\lambda(z) - \lambda(0)) \\
  &=& \arg \left[\langle \nu_{e}(0)| \nu_{e}(z)\rangle  \right]  - \Im \int_{0}^{z}
  \braket{ \nu_{e}(z^{\prime})|\dot{\nu}_{e}(z^{\prime})} d z^{\prime}  \ ,
\end{eqnarray}
showing that a term proportional to the identity in the wolfenstein effective Hamiltonian has no effect on the geometric invariants, contrarily to what affirmed in the reference \cite{Lu2021}.


In addition to $\Phi^{g }_{\nu_{e}}(z)$ and $\Phi^{g }_{\nu_{\mu}}(z)$, one can also define two geometric invariants associated to the neutrino oscillation between different flavors, i.e.
 \begin{eqnarray} \label{fasemix1}
\Phi^{g }_{\nu_{e}\rightarrow \nu_{\mu}}(z)
\!\!& = &\!\!
  - \Im \int_{0}^{z}  \langle  \nu_{e}(z^{\prime})| \dot{\nu}_{\mu}(z^{\prime})\rangle  d z^{\prime}\, ,
\\  \label{fasemix2}
\Phi^{g }_{\nu_{\mu}\rightarrow \nu_{e}}(z)
\!\!& = &\!\!
  - \Im \int_{0}^{z}  \langle  \nu_{\mu}(z^{\prime})| \dot{\nu}_{e}(z^{\prime})\rangle  d z^{\prime}\, .
 \end{eqnarray}
Explicitly, one has
\begin{eqnarray}
\label{fasemix1aa}
\Phi^{g }_{\nu_{e}\rightarrow \nu_{\mu}}(z)   =  \Phi^{g }_{\nu_{\mu}\rightarrow \nu_{e}}(z)  =  \left(\frac{\Delta m_{m}^{2}}{4 E}\,\sin 2\theta_{m}\; \cos \phi\; \right) z\, ,
\end{eqnarray}
Also in this last evaluation we have neglected the constant term proportional to the identity operator in the neutrino Hamiltonian $H_0 = \left( \frac{m_1^2 + m_2^2}{4E} + \frac{G_F n_e}{\sqrt{2}} \right) \mathbb{I}$ and we have taken into account that for any time $z$, we have $\braket{\nu_{e}(z)| \nu_{\mu}(z)} = 0$. It is straightforward to prove that also the geometric invariant of Eq. \eqref{fasemix1} is not affected by this additional term in the Hamiltonian:
\begin{eqnarray} \label{GaugeInvariance}
 \nonumber \Phi^{g }_{\nu_{e}\rightarrow \nu_{\mu}}(z)
\!\!& = &\!\!
  - \Im \int_{0}^{z} e^{- i \lambda(z')} \braket{ \nu_{e}(z^{\prime})| \frac{d}{dz'}  \left(e^{i \lambda (z')}\nu_{\mu}(z^{\prime}) \right)}  d z^{\prime} \\
  \nonumber &=& - \Im \int_{0}^{z} i \lambda'(z')  \braket{ \nu_{e}(z^{\prime})| \nu_{\mu}(z^{\prime})} dz' - \Im \int_{0}^z \braket{ \nu_{e}(z^{\prime})| \dot{\nu}_{\mu}(z^{\prime})} dz' \\
  &=& - \Im \int_{0}^z \braket{  \nu_{e}(z^{\prime})| \dot{\nu}_{\mu}(z^{\prime})} dz' \ .
\end{eqnarray}
The trivial calculation above also proves that $\Phi^{g }_{\nu_{e}\rightarrow \nu_{\mu}}(z)$ as defined in eq. \eqref{fasemix1} is indeed gauge invariant, since eq. \eqref{GaugeInvariance} actually holds for any smooth function $\lambda(z)$. It is also easy to see that the Eqs. \eqref{fasemix1} are reparametrization invariant.
Notice that the original full Mukunda-Simon definition (e.g. Eq.~\eqref{fase1}) fails to be gauge invariant when the distinct neutrino states $\ket{\nu_e (z)}, \ket{\nu_{\mu}(z)}$ are used as $\arg \left[\braket{\nu_{e}(0)| \nu_{\mu}(z)}\right] - \Im \int_{0}^{z}
 \braket{\nu_{e}(z^{\prime})| \dot{\nu}_{\mu}(z^{\prime})}  d z^{\prime}$  .
\

This could be expected, because the geometric phase is defined on a single trajectory in Hilbert space and not for two distinct paths in Hilbert space. The quantity defined in equation \eqref{fasemix1} is indeed not a geometric phase, but a geometric invariant enjoying the same properties as the geometric phase, and therefore in principle can be detected.

Moreover, differently from $\Phi^{g }_{\nu_{e}}(z)$ and $\Phi^{g }_{\nu_{\mu}}(z)$ the two geometric invariants introduced in eqs.~\eqref{fasemix1} and~\eqref{fasemix2} depend on the Majorana phase $\phi$.
In the case of a Dirac neutrino, the phase $\phi$ can be removed by means of a rephasing of the neutrino fields and the invariants of Eqs.(\ref{fasemix1aa}) reduce to
\begin{eqnarray} \label{fasemixD}\nonumber
\Phi^{g }_{\nu_{e}\rightarrow \nu_{\mu}}(z) & = & \Phi^{g }_{\nu_{\mu}\rightarrow \nu_{e}}(z)=  \left(\frac{\Delta m_{m}^{2}}{4 E}\,\sin 2\theta_{m}\;  \right) z .
\end{eqnarray}
Notice that the same result is obtained if one uses the mixing matrix $U^{(1)}$ for the definition of the flavor states.
We point out that for the discussion in the previous section, in the case of Majorana neutrinos, such a dependence cannot be removed by a lepton field rephasing transformation,
as affirmed in Ref.~\cite{Lu2021}.
Then the two geometric invariants defined in  Eqs.~(\ref{fasemix1}) and~(\ref{fasemix2}) depend on the Majorana CP-violating phase, and as well-defined geometric quantities, they are in principle measurable.

\section{Additional remarks on the geometric phase}

 In this section we analyze the kinematic and the geometric approach to the geometric phase. We define a new off--diagonal geometric invariant $\tilde{\phi}^{g}$ (a $U(1)$ gauge invariant and reparametrization invariant quantity). We show that the Majorana phase can be relevant also on the projective Hilbert space, therefore entering geometric quantities and we comment on the possibility  to reveal the Majorana phase in interference experiments.

 \subsection{Kinematic Approach}

 In the kinematic approach the geometric phase for a quantum system, whose state vector $\ket{\psi (s)}$ belongs to an open curve $\Gamma$ of unit vectors in Hilbert space, for $s \in [s_1,s_2]$, is defined as the difference between the total and dynamic phase:  
 \begin{eqnarray}
   \phi^{g} (\Gamma)  =  \phi^{(total)}(\Gamma) - \phi^{(dynamical)} (\Gamma) =   \mathrm{arg} \langle \psi(s_1) \ket{\psi(s_2)} - \Im \int_{s_1}^{s_2} ds \langle \psi (s) \ket{\dot{\psi}(s)} \ . 
 \end{eqnarray}
Let us consider a possible generalization of this definition to neutrino transitions between different flavors.
Let us start by considering the scalar products $\bra{\nu_{\alpha}(0)} \nu_{\beta}(z) \rangle$ and  for $\alpha,\beta = e, \mu$, whose argument represents the total phase in the kinematic approach.
The product is independent on the Majorana phase for $\alpha = \beta$ and also for the off-diagonal phases $\alpha \neq \beta$ when the mixing matrix is chosen as $U^{(1)}$ in eq. \eqref{MixingMatrices}.

However, it is straightforward to see that if the mixing matrix is $U^{(2)}$, the off-diagonal total phases depend on the Majorana phase $\phi$. To show this, let us write the flavor states according to the mixing matrix $U^{(2)}$:
\begin{eqnarray*}
 \ket{\nu_{e} (z)} &=& \cos \theta e^{i \frac{\Delta m^2}{4 E}z} \ket{\nu_1} + e^{i \phi} \sin \theta e^{-i \frac{\Delta m^2}{4 E}z} \ket{\nu_2} \\
 \ket{\nu_{\mu} (z)} &=& - e^{-i\phi}\sin \theta e^{i \frac{\Delta m^2}{4 E}z} \ket{\nu_1} +  \cos \theta e^{-i \frac{\Delta m^2}{4 E}z} \ket{\nu_2}  \ .
\end{eqnarray*}
The off-diagonal scalar products read
\begin{eqnarray}\label{ScalarProducts}
 \nonumber \bra{\nu_{e}(0)} \nu_{\mu}(z) \rangle &=& e^{-i\phi} \cos \theta \sin \theta \left( e^{-i \frac{\Delta m^2}{4 E}z} - e^{i \frac{\Delta m^2}{4 E}z}\right) \\
  \nonumber \bra{\nu_{\mu}(0)} \nu_{e}(z) \rangle &=& e^{i\phi} \cos \theta \sin \theta \left( e^{-i \frac{\Delta m^2}{4 E}z} - e^{i \frac{\Delta m^2}{4 E}z}\right)
\end{eqnarray}
which evidently do depend on $\phi$.
Therefore, as far as the scalar products in eqs. \eqref{ScalarProducts} are concerned, the two mixing matrices $U^{(1)}$ and $U^{(2)}$ are not equivalent. In particular, the dependence on the Majorana phase $\phi$, which shows up for $U^{(2)}$ cannot be removed by means of the  charged lepton field rephasing, as discussed in section 2.

If we try to implement the kinematic definition for the off-diagonal case straight away, namely by putting
\begin{equation}
 \Phi^{g}_{\alpha, \beta} (z) = \arg \left[\bra{\nu_{\alpha}(0)} \nu_{\beta} (z) \rangle \right] - \Im \int_0^{z} dz' \bra{\nu_{\alpha} (z')} \dot{\nu}_{\beta} (z') \rangle
\end{equation}
we obtain an object which is not geometrical, in the sense that the above quantity is not $U(1)$ gauge invariant. Gauge invariance can be restored by replacing the last term with one of its diagonal counterparts:
\begin{equation}\label{fasemix3}
 \tilde{\Phi}^{g}_{\alpha, \beta} (z) = \arg \left[\bra{\nu_{\alpha}(0)} \nu_{\beta} (z) \rangle \right] - \Im \int_0^{z} dz' \bra{\nu_{\alpha(\beta)} (z')} \dot{\nu}_{\alpha (\beta)} (z') \rangle \ .
\end{equation}
Just like eqs. \eqref{fasemix1} and \eqref{fasemix2} one can easily prove that eqs. \eqref{fasemix3} are $U(1)$ gauge invariant and that they depend explicitly on the Majorana phase $\phi$ if one considers the mixing matrix $U^{(2)}$.

Both the definitions \eqref{fasemix1}, \eqref{fasemix2} and \eqref{fasemix3} are geometric invariants, defined in the kinematic approach, employable for neutrino flavor transitions. They are economical, in the sense that only a minimal modification is made to the usual definition of the geometric phase in order to preserve its essential features. Indeed, they are geometric invariants in the usual meaning, as they are $U(1)$ gauge invariant and reparametrization invariant. And, most importantly, they show an explicit dependence on the Majorana phase $\phi$.
Whether the quantities defined above are observable or not is a completely different issue, which we make no claim about.

There exists another definition of off-diagonal geometric phase, due to 
Manini and Pistolesi \cite{Manini}. This phase, when used for neutrino states, is independent on the Majorana phase \cite{Johns2021}. Anyway the so defined geometric phase is by construction a higher order gauge invariant (it is invariant under $U(1) \times U(1)$ for two flavors and under $\left[ U(1) \right]^n$ in the general $n$ flavor case) and it is obvious that it cannot depend on the Majorana phase. This is actually true of any $ U(1) \times U(1)$ invariant, since in their evaluation one can always individually rephase the states so to eliminate the Majorana phase $\phi$.
While the off-diagonal phase \cite{Manini} is certainly well-defined, it is not at all clear, in the case of Majorana neutrinos, whether this invariant catches all the relevant geometric information.
We leave this issue open, with the awareness that the quantum mechanical description of Majorana neutrinos is somewhat ambiguous. After all, if a more fundamental quantum field theoretical description of neutrinos must be considered, its quantum mechanical counterpart should reproduce all the relevant physics as closely as possible. In this respect, the Dirac neutrino fields are described by a $U(1)$ gauge invariant Lagrangian, meaning that the phase of the field is pure gauge. It is then reasonable that their quantum mechanical analogues, the Dirac neutrino states, preserve the rephasing freedom.

On the contrary, the Majorana Lagrangian is not $U(1)$ invariant, so that it is not possible to freely rephase the Majorana field. We believe that the corresponding quantum mechanical theory should somehow take this lack of rephasing freedom into account.

\subsection{Geometric Approach}

Let us discuss now the possible role of the Majorana phase in the projective Hilbert space for neutrinos. We consider
the neutrino state  parametrized as \cite{Johns2021}
\begin{equation}\label{JonPar}
 \ket{\nu} = e^{i \chi} \begin{pmatrix}
                         \cos \frac{\theta}{2} e^{i \frac{\phi}{2}} \\   \sin \frac{\theta}{2} e^{-i \frac{\phi}{2}}
                        \end{pmatrix}
\end{equation}
where $\theta$ and $\phi$ are the polar angles on the Bloch sphere and $\chi$ is a phase factor. Notice that the phase factor $\chi$ cannot be chosen arbitrarily if the neutrino state is of the Majorana kind. The reason is that the Majorana neutrino states cannot be rephased freely. Indeed, if one could arbitrarily choose choose $\chi$ in order to remove the Majorana phase, one would then effectively have a Dirac neutrino state, which cannot be used for Majorana neutrinos

Notice that the density matrix for flavor neutrinos $\rho_{\sigma} = \ket{\nu_{\sigma}} \bra{\nu_{\sigma}}$, for $\sigma = e, \mu$, does not depend on the phase factor $\chi$, if the parametrization \eqref{JonPar} is referred to the flavor state $\ket{\nu_{\sigma}}$, since $\chi$ is a global phase factor. 

On the other hand if one considers the parametrization \eqref{JonPar} as referred to the individual mass states $\ket{\nu_j}$, $j=1,2$, the density matrix for flavor neutrinos depends on the phase factors $\chi_j$. In order to show this, we consider for simplicity the two flavor neutrino case, with the two mass states parametrized as
\begin{equation}
 \ket{\nu_1} = e^{i \chi_1} \begin{pmatrix}
                         \cos \frac{\theta_1}{2} e^{i \frac{\phi_1}{2}} \\   \sin \frac{\theta_1}{2} e^{-i \frac{\phi_1}{2}}
                        \end{pmatrix}  \ \ \ \ \ \ \ \ \
                         \ket{\nu_2} = e^{i \chi_2} \begin{pmatrix}
                         \cos \frac{\theta_2}{2} e^{i \frac{\phi_2}{2}} \\   \sin \frac{\theta_2}{2} e^{-i \frac{\phi_2}{2}}
                        \end{pmatrix}
\end{equation}
The polar angles $\theta_j$ and $\phi_j$ are such that $\ket{\nu_1}$ and $\ket{\nu_2}$ are orthogonal to each other. Taking the $U^{(2)}$ mixing matrix, the electron neutrino state is then
\begin{equation}
 \ket{\nu_e} = \cos \theta e^{i \chi_1} \begin{pmatrix}
                         \cos \frac{\theta_1}{2} e^{i \frac{\phi_1}{2}} \\   \sin \frac{\theta_1}{2} e^{-i \frac{\phi_1}{2}}
                        \end{pmatrix}  + \sin \theta e^{i (\phi + \chi_2)}\begin{pmatrix}
                         \cos \frac{\theta_2}{2} e^{i \frac{\phi_2}{2}} \\   \sin \frac{\theta_2}{2} e^{-i \frac{\phi_2}{2}}
                        \end{pmatrix}   =  \cos \theta e^{i \chi_1} \ket{\bar{\nu}_1} + \sin \theta e^{i(\phi + \chi_2)} \ket{\bar{\nu}_2}
\end{equation}
where $\ket{\bar{\nu}_j}$ are the mass states without the phase factors $\chi_j$. Then the associated density matrix is
\begin{equation}
 \rho_e = \ket{\nu_e} \bra{\nu_e} = \cos^2 \theta \ket{\bar{\nu}_1} \bra{\bar{\nu}_1} + \cos \theta \sin \theta \left(e^{i(\chi_1 - \phi - \chi_2)} \ket{\bar{\nu}_1} \bra{\bar{\nu}_2} + e^{i(\phi + \chi_2 - \chi_1)} \ket{\bar{\nu}_2} \bra{\bar{\nu}_1} \right) + \sin^2 \theta \ket{\bar{\nu}_2} \bra{\bar{\nu}_2}
 \end{equation}
which evidently depends on the phases $\chi_1,\chi_2, \phi$. $\chi_1,\chi_2, \phi$ are then relevant also for the projective Hilbert space. A similar result is obtained for different choices of the mixing matrix.

Following the reference \cite{Johns2021} we work in the mass basis in order to consider the evolution of the flavor states. For simplicity we consider two flavors, but the argument can be easily generalized to three flavors. The two flavor analogues of eqs. (32) and (33) of the reference \cite{Johns2021} are respectively
\begin{equation}
 \ket{\nu_{\beta}} = \bar{U}^*_{\beta 1 } \ket{\nu_1} + \bar{U}^*_{\beta 2} e^{- i \phi } \ket{\nu_2}
\end{equation}
with $\beta = e, \mu$ and
\begin{equation}
 \ket{\nu_{\beta}} = \bar{U}^*_{\beta 1 }e^{i \chi_1} \begin{pmatrix}
                         \cos \frac{\theta_1}{2} e^{i \frac{\phi_1}{2}} \\   \sin \frac{\theta_1}{2} e^{-i \frac{\phi_1}{2}}
                        \end{pmatrix} + \bar{U}^*_{\beta 2} e^{- i \phi } e^{i \chi_2}\begin{pmatrix}
                         \cos \frac{\theta_2}{2} e^{i \frac{\phi_2}{2}} \\   \sin \frac{\theta_2}{2} e^{-i \frac{\phi_2}{2}}
                        \end{pmatrix} \ .
\end{equation}
Here $\bar{U}^*_{\beta j}$ represent the PNMS matrix elements without the Majorana phase. Notice that the above equations imply that the mixing matrix is chosen as $U^{(1)}$, for which the Majorana phase is attached to a single mass state. Assuming, for the sake of the argument, that one could freely choose $\chi_1$ and $\chi_2$, then one could clearly remove the Majorana phase $\phi$ from all the flavor states. However. as we have discussed, the mixing matrix $U^{(1)}$ is equivalent to the mixing matrix without Majorana phases. Let us consider now the mixing matrix $U^{(2)}$, for which the flavor states are
\begin{equation}
 \ket{\nu_e} = \bar{U}^*_{\beta 1 }e^{i \chi_1} \begin{pmatrix}
                         \cos \frac{\theta_1}{2} e^{i \frac{\phi_1}{2}} \\   \sin \frac{\theta_1}{2} e^{-i \frac{\phi_1}{2}}
                        \end{pmatrix} + \bar{U}^*_{\beta 2} e^{- i \phi } e^{i \chi_2}\begin{pmatrix}
                         \cos \frac{\theta_2}{2} e^{i \frac{\phi_2}{2}} \\   \sin \frac{\theta_2}{2} e^{-i \frac{\phi_2}{2}}
                        \end{pmatrix}
\end{equation}

\begin{equation}
 \ket{\nu_{\mu}} = \bar{U}^*_{\beta 1 }e^{i \phi} e^{i \chi_1} \begin{pmatrix}
                         \cos \frac{\theta_1}{2} e^{i \frac{\phi_1}{2}} \\   \sin \frac{\theta_1}{2} e^{-i \frac{\phi_1}{2}}
                        \end{pmatrix} + \bar{U}^*_{\beta 2}  e^{i \chi_2}\begin{pmatrix}
                         \cos \frac{\theta_2}{2} e^{i \frac{\phi_2}{2}} \\   \sin \frac{\theta_2}{2} e^{-i \frac{\phi_2}{2}}
                        \end{pmatrix}  \ .
\end{equation}
In this case, the Majorana phase cannot be removed even if $\chi_1$ and $\chi_2$ can be chosen freely. Indeed, setting $\chi_2 = \phi = - \chi_1$, in order to eliminate the Majorana phase $\phi$, one obtains the following flavor states:
\begin{equation}
 \ket{\nu_e} = \bar{U}^*_{\beta 1 } e^{-i \phi} \begin{pmatrix}
                         \cos \frac{\theta_1}{2} e^{i \frac{\phi_1}{2}} \\   \sin \frac{\theta_1}{2} e^{-i \frac{\phi_1}{2}}
                        \end{pmatrix} + \bar{U}^*_{\beta 2} \begin{pmatrix}
                         \cos \frac{\theta_2}{2} e^{i \frac{\phi_2}{2}} \\   \sin \frac{\theta_2}{2} e^{-i \frac{\phi_2}{2}}
                        \end{pmatrix}
\end{equation}

\begin{equation}
 \ket{\nu_{\mu}} = \bar{U}^*_{\beta 1 }\begin{pmatrix}
                         \cos \frac{\theta_1}{2} e^{i \frac{\phi_1}{2}} \\   \sin \frac{\theta_1}{2} e^{-i \frac{\phi_1}{2}}
                        \end{pmatrix} + \bar{U}^*_{\beta 2}  e^{i \phi}\begin{pmatrix}
                         \cos \frac{\theta_2}{2} e^{i \frac{\phi_2}{2}} \\   \sin \frac{\theta_2}{2} e^{-i \frac{\phi_2}{2}}
                        \end{pmatrix}  \ .
\end{equation}
Here, the Majorana phase $\phi$, clearly has not disappeared and is not a global phase factor. Other choices for $\chi_1$ and $\chi_2$ are possible, for instance $\chi_2 = - \chi_1 = \frac{\phi}{2}$. With this choice the flavor states acquire two distinct phase factors ($e^{- i \frac{\phi}{2}}$ for $\ket{\nu_e}$ and $e^{ i \frac{\phi}{2}}$ for $\ket{\nu_{\mu}}$. Even in this case, the Majorana phase survives as a relative phase between $\ket{\nu_e}$ and $\ket{\nu_{\mu}}$. In any case we point out that the $\chi$ parameters cannot be chosen arbitrarily due to the Majorana nature of the states.
From what we have shown, it follows that the Majorana phases, like the $\chi$ parameters, are geometrically relevant in the Projective Hilbert space and then they can show up in geometric phases.

In light of the above discussion, the Majorana phases are potentially detectable. Indeed, if one could make a measurement of the quantity
\begin{equation}
 |\bra{\nu_e}\left(A_1 \ket{\nu_1} + A_2 \ket{\nu_2} + A_3 \ket{\nu_3}\right)|^2 = |A_1 \bar{U}_{e1} e^{i \alpha_1} + A_2 \bar{U}_{e2} e^{i \alpha_2} + A_3\bar{U}_{e3} |^2
\end{equation}
for arbitrary $A_1, A_2, A_3$, then one could access the Majorana phases.
We point out that the Majorana phases are in principle observable in oscillation experiments since, as shown, they can appear in the evolution of the flavor neutrino states, independently of how the mixing matrix is factorized \footnote{Another quantity that one can consider is the quantum geometric tensor, defined as
\begin{equation}
 T_{\mu \nu} (\psi) = \bra{\partial_{\mu}\psi} \partial_{\nu} \psi \rangle - \bra{\partial_{\mu} \psi} \psi \rangle \bra{\psi} \partial_{\nu} \psi \rangle .
\end{equation}
We notice that in the diagonal case (as $T_{\mu \nu} (\ket{\nu_e})$ and $T_{\mu \nu} (\ket{\nu_{\mu}})$) the geometric tensor is a higher order ($U(1)^n$) gauge invariant and therefore does not depend on the Majorana phases. The same considerations made about the off-diagonal geometric phase apply to the so defined geometric tensor about its relevance for Majorana neutrino states.}.

We remark that nowhere we have claimed that the Majorana phases are geometric in nature. However what we have shown is that there exist some geometric invariants, like those defined above, that bear a dependence on the Majorana phases. In other words, despite the Majorana phases not being geometric by themselves, they can be geometrically relevant.
Again we stress that Majorana neutrino states cannot be rephased arbitrarily, so there is no $U(1)^n$ gauge invariance in the description of Majorana neutrinos.

\subsection{Interference vs. oscillation experiments}

As a last comment, consider the probability \cite{Johns2021}
\begin{equation}
 P_{\nu \rightarrow \nu_e} (z) = |\bra{\nu_e} \nu (z) \rangle|^2= \left|\sum_{i,j} \bra{\nu_j} U_{ej} \left(A_e U^*_{ei} + A_{\mu} U^*_{\mu i} + A_{\tau} U^*_{\tau i} \right)e^{-iE_iz} \ket{\nu_i} \right|^2 \ ,
\end{equation}
with $A_{e,\mu,\tau}$ arbitrary. By using the mixing matrix $U^{(2)}$ one has $P_{\nu \rightarrow \nu_e} (z)$ depending on the Majorana phase. Therefore, in principle, the Majorana phase could be observed in interference experiments. Indeed, even if neutrinos are produced and detected as defined flavor states (with only one of the $A_{\mu}$ coefficients non zero), in the papers \cite{Capolupo2019} it has been shown that in the presence of decoherence, which may be present in long baseline experiments, the oscillation formulas can depend on the Majorana phase.
Hence interference and oscillation experiments can in principle reveal the Majorana phases.

\section{Conclusions}

We discussed the geometric phase for neutrinos and  the possibility to distinguish,  through it,
the Majorana neutrinos from those of the Dirac type. 
We defined geometric invariants which are $U(1)$ gauge and reparametrization invariant.

We demonstrated that the geometric invariants, associated to transitions between different neutrino flavors, may depend on the representation of the mixing matrix in the case of Majorana neutrinos. Consequently, we have shown that they are sensitive to the fundamental nature of neutrinos. In particular, geometric phases depending on the Majorana phase are obtained by considering the $U^{(2)}$ mixing matrix,
while the use of $U^{(1)}$ matrix  leads to geometric phases independent on the Majorana phase.
We have proven that the appearance of this phase cannot be explained, nor eliminated, by means of a simple lepton rephasing transformation.
We also described a gedanken experiment, showing that the redefinition of the mixing matrix due to the
rephasing, modifies some physical quantities.
We considered the kinematic and the geometric approach to the geometric phase and we demonstrated that the Majorana phases can be relevant in  geometric invariants.

In conclusion, geometric invariants can represent theoretical tools  to distinguish between
Dirac and Majorana neutrinos.

\appendix

\section{Geometric Phase and Wavefunction Collapse}

In this appendix we comment on the recent result on the geometric phase of ref.~\cite{Lu2021}. There the contribution to the geometric invariants, due to the collapse of the wavefunction induced by the projective measurement operation, has been considered. The collapse of the wavefunction is taken into account through an estimate in which the projective measurement process is carried out through a process in which, in a time $ \tau $, the state at time $z-\tau$ is led into one of the two flavor eigenstates.
In the end, the time $ \tau $ is then made to tend to zero to recover the unitarity of the measurement process, a property considered central by the author of the ref.~\cite{Lu2021}.

This approach is flawed in at least two respects: from the conceptual point of view, the introduction of the wavefunction collapse is not justified and leads to incorrect conclusions; from the technical point of view, the description of the collapse adopted is arbitrary and incorrect. The latter, in particular, gives an indication of the approximation with which the comment was written.

Let us focus on the first point. In order to define the geometric invariants associated to the mixing, namely Eqs. \eqref{fasemix1} and \eqref{fasemix2}, it is not necessary that the transition $\ket{\nu_e} \rightarrow \ket{\nu_{\mu}}$ takes places, but it is sufficient to consider the evolved state $\ket{\nu_{\mu} (t)}$ which is by itself a combination of the initial flavor eigenstates. Asserting that the neutrino state at time $t$ is $\ket{\nu_{\mu}(t)}$ is of course different from asserting that the neutrino state at time $t$ is $\ket{\nu_{\mu}(0)}$. The last assertion, in particular, implies a non-unitary evolution of the neutrinos at some point.
Moreover, even though it is not clear how to detect the geometric invariants of Eqs. \eqref{fasemix1} and \eqref{fasemix2}, this fact does not imply that they are not observable in principle.

To justify the introduction of the wavefunction collapse at least one of the following hypotheses should be true:
\begin{enumerate}
 \item The detection relies on a projective measure.
 \item  The collapse of the wave function is intrinsically associated with the physical phenomenon being analyzed.
\end{enumerate}

The second assumption is certainly wrong. Indeed the oscillation between two levels is a well-known phenomenon in several fields of quantum mechanics such as, to provide an example in a very different field, the vacuum Rabi oscillation~\cite{Loundon2000}, and never involve a collapse of the wave function process. On the other hand, we make no claim about how a possible experimental detection of these geometrical invariants is to be performed.

In addition, in Ref.~\cite{Lu2021} it is ignored the fact that not all the possible measures that can be done on a quantum system must include demolition of the quantum state~\cite{Lupescu2007}.
The design of a non-demolition measure for neutrinos would be a difficult task to achieve, but if one wants to prove that a quantity cannot be observed, one must consider all the possible ways to realize a measurement.

Even assuming the necessity of including the wavefunction collapse, for whatever reason, the definition of the geometric invariant presented in~\cite{Lu2021} is not correct.
Indeed, the wavefunction collapse is taken into account considering an arbitrary non-unitary evolution, for which unitarity is restored in the limit of a vanishing collapse time $\tau$ (see eq. (14) of the ref. \cite{Lu2021}). In any case, while it can be assumed that  the wavefunction collapse is very rapid, one cannot expect it to be a unitary process. In fact, it is well known that the measurement of an observable on a pure state which is not an eigenstate of the corresponding operator produces a mixed state with reduced purity ~\cite{Nielsen2000}. This general fact implies that the wavefunction collapse cannot be represented by a unitary operator preserving the purity of the state.

A coherent analysis of the wavefunction collapse process, should certainly start from considering, during the measurement process, the neutrino as an open system which interacts with the experimental device.
The dynamics of the neutrino during the measurement process must be described by a Gorini-Kossakowski-Sudarshan-Lidbland master equation~\cite{Lindblad1976,Kossakowski1976}
\begin{equation}
\frac{\partial \rho(t)}{\partial t}=-\frac{\imath}{\hbar}[H,\rho(t)]+D[\rho(t)] \ .
\end{equation}
Here $\rho(t)$ is the time-dependent density matrix of the neutrino while the non unitary part of the evolution is described by the dissipator $D[\rho(t)]$ defined as
\begin{equation}
D[\rho(t)]=\frac{1}{2}\sum_{i=1}^{N-1} a_{ij}\left([F_i \rho(t),F_j^\dagger]+[F_i, \rho(t) F_j^\dagger] \right)\ .
\end{equation}
The coefficients $a_{ij}$ of the Kossakowski matrix, can be derived by the properties of the measurement device~\cite{Benatti2000, Benatti2001} while $F_i$
are a set of traceless operators such that $Tr(F_i^\dagger F_j)=\delta_{ij}$ .
In the two flavor neutrino mixing $F_i$ can be represented by the Pauli matrices, and in the three flavor generalization the $F_i$ can be represented by the Gell-Mann matrices $\lambda_i$.
Therefore the eq. (14) of ref. \cite{Lu2021} and the following eqs. are flawed and not physically consistent.

\section*{Acknowledgements}
A.C., G.L. and A.Q. acknowledge partial financial support from MUR and INFN, A.C. and G.L. also acknowledge the COST Action CA1511 Cosmology
and Astrophysics Network for Theoretical Advances and Training
Actions (CANTATA),
S.M.G acknowledges support from the Croatian Science Funds Project No.
IP-2019–4–3321 and
the QuantiXLie Center of Excellence, a project co–financed by the
Croatian Government and European
Union through the European Regional Development Fund–-the
Competitiveness and Cohesion Operational Programme (Grant
KK.01.1.1.01.0004).
B.C.H. acknowledges the support of the Austrian Science Fund (FWFP26783).

\end{document}